\documentclass[12pt,titlepage,a4paper]{article}
\usepackage{fancyheadings}
\usepackage{epsfig}
\usepackage{amssymb}
\usepackage{amsmath}
\usepackage{amsfonts}
\usepackage{color,hyperref}
\usepackage{color,graphicx}

\title {\sc {\bf Cryptanalysis \\of a \\Chaotic Image Encryption Algorithm}}
\author{Nikhil Balaji \\ Department of Electronics and Communicaton Engineering\\ National Institute of Technology Karnataka, Surathkal\\ {\bf nikhilrameshbalaji@gmail.com} \\ \\  Nithin Nagaraj \\School of Natural and Engineering Sciences \\ National Institute of Advanced Studies\\ IISc. Campus, Bangalore 560012 \\ {\bf nithin@nias.iisc.ernet.in}}
\date {January 10, 2008}

\begin{document}
\maketitle

\begin{abstract}
Line map, an invertible, two-dimensional chaotic encryption
algorithm was introduced recently. In this paper, we propose several
weaknesses of the method based on standard cryptanalytic attacks. We
perform a side-channel attack by observing the execution time of the
encryption algorithm and successfully reduce the key space by a
factor of $10^{4}$ for a key length of 16 digits. We find the
existence of equivalent keys which reduce the key space by a
significant margin, even in the absence of any attack. Also, we find
that the ciphertext is not sensitive to small changes in the
plaintext due to poor diffusion.
\end{abstract}

\section{Introduction}

The need for fast and secure encryption schemes for bulky data such
as audio, images and video has prompted the advance of chaos-based
encryption schemes over traditional schemes, which make use of the
inability to perform factorization of large numbers (RSA) or to
solve the discrete logarithm problem (El Gamal) in a fast and
efficient manner. Recent advances like distributed computing,
quantum computing and new advances in algorithmic number theory can
potentially make such traditional schemes totally unusable in the
future. In contrast, chaos-based encryption schemes are fast and
easily realized in both hardware and software, which makes it more
suitable for data encryption.

The two basic properties of chaotic systems are the sensitivity to
initial conditions and mixing. Sensitivity to initial conditions
means that when a chaotic map is iteratively applied to two
initially close points, the iterates quickly diverge, and bear no
correlation after a few iterations. Sensitivity to parameters causes
the properties of the map to change quickly when the parameters on
which the map depends are mildly disturbed. Mixing is the tendency
of the system to quickly confuse small portions of the state space
into an intricate network so that two nearby points in the system
totally lose the correlation they once shared and get scattered all
over the state space. These properties are the key aspects of
chaotic maps that have allowed them to be used to generate
complicated patterns of pixels and the gray levels in an image.

Traditionally, there are two  ways in which chaos is used in image
encryption schemes: 1) 1-D chaotic maps like logistic map and
generalizations of logistic map~\cite{Mahesh2007} to generate
pseudo-random bits with desired statistical properties to realize
secret encryption operations.  2) 2-D chaotic maps like Arnold's cat
map, Baker map or fractal-like curves to realize secret permutations
of digital images. The first approach has been widely used to design
chaotic stream ciphers, while the second is specially employed by
chaos-based block encryption schemes. (Refer Table 1 for a brief
overview on some recent chaos-based image encryption schemes and
their properties).

The rest of the paper is organized as follows: The next section
gives a brief description of the original line map algorithm and its
use in encrypting and decrypting images as developed
in~\cite{Feng2006}. Section 3 points out to the major weaknesses of
the algorithm and gives a list of attacks that can be performed on
the scheme owing to the weaknesses discussed, Section 4 introduces
the Joint Entropy function as a plausible metric for quantifying how
well an algorithm or chaotic map, makes the encrypted image look
random and Section 5 concludes the paper.

\begin{table}
\centering \caption{A brief survey of the chaotic image encryption
literature}
\begin{tabular}{|p{4cm}|p{4.5cm}|p{5.5cm}|}
%\caption{\textbf{A Brief Survey of the Chaotic Image encryption literature}}
\hline
\textbf{Scheme} & \textbf{Strengths} & \textbf{Weakness and known attacks}\\

\hline

Scharinger J ``Fast encryption of image data using chaotic
Kolmogorov Flows''~\cite{Scharinger1998}
 & Unstable dynamics of the kolmogorov flow ensures good mixing.
& Size of key space depends on size of image. \\

\hline
 Fridrich J  ``Symmetric ciphers based on two-dimensional chaotic maps''~\cite{Fridrich1997}~\cite{Fridrich1998}
& Cat map: high parameter sensitivity; Baker and Standard map: Large key space
&Catmap: small key space; Baker and Standard map: Higher computational complexity.~\cite{Lian2005}\\
\hline
Yen et al. ``Bit Recirculation Image Encryption" (BRIE) ~\cite{Yen1999} ~\cite{Yen2002}
&Low hardware and computational complexity, high encryption speed, no distortion
&Weak keys, vulnerable to known/chosen-plaintext attacks, since a mask image that is equivalent to secret key can be derived from a pair of plain-image and cipher-image.~\cite{Li2002}\\
\hline Yen et al. ``Chaos key  based algorithm" (CKBA)
~\cite{Yen2000} &Low computational complexity, Good mixing/confusion
in pixels &Weak to chosen and known plaintext attacks, security to
brute-force attack is also questionable.~\cite{Shujun2002}\\ \hline
Yen et al. ``Hierarchical Chaotic Image Encryption"(HCIE)
~\cite{Yen1998} ~\cite{Guo1999} ~\cite{ChengYen2000} &Parallel and
pipelined realization of the scheme is possible, hence high
encryption speed & Permutation-only scheme. Cannot resist chosen plaintext attack.~\cite{SLi2004}\\
\hline
Mao et al. ``Symmetric image encryption based on 3D cat maps"~\cite{Mao2004}
&Large key space, good key sensitivity and resistance to statistical and differential attacks.
& Insensitivity to changes in plaintext and key stream generated by any key,poor diffusion function. ~\cite{Kai2005} ~\cite{Cli2007}\\
\hline
Mao et al. ``Fast image encryption scheme based on the 3D chaotic baker maps"~\cite{Mao2003}
&Large key space
&Insensitivity to changes in plaintext and key stream generated by any key,poor diffusion function.~\cite{Cli2007}\\
\hline

\end{tabular}
\end{table}

\section{Line map algorithm}
The line map is an invertible, two-dimensional chaotic mapping
introduced by Feng~\cite{Feng2006}. It achieves lossless encryption
with a good sensitivity to change in key and exhibits good mixing
properties. The line map proposed has two forms namely the left line
map and the right line map. Here each pixel of a line perpendicular
to the diagonal is inserted between the adjacent two pixels of the
next line perpendicular to the diagonal. It is suggested that to
arrange the original image in to a row of pixels is a process of
{\it stretching} and rearranging the row of pixels to image is an
operation of {\it folding}. From the point of view of these
stretching and folding operations, the line map is chaotic, similar
to the Baker map or the Cat map.

Let $A(i,j)$ be a square image matrix; $l(i), i=1,2,\dots,N.N $is a
one dimensional vector mapped from $A(i,j)$ and $fix(X)$ is a
function which rounds the elements of $X$ to the nearest integers
towards zero.The algorithm for the line map for square images is
given by:

\begin{equation*}
l\left(\sum_{k=0}^{i-1}(4k-1)+j+1\right)\ =A\left(fix\frac{(4i-j+1)}{2},fix\frac{(j+1)}{2}\right)
\end{equation*}
\begin{center}
$j=1,2,\dots,4i-1$\\
$i=1,2,\dots,fix(\frac{N}{2})$\\
\end{center}
\begin{equation*}
l\left(\sum_{k=0}^{fix(\frac{N}{2})}(4k-1)+\sum_{k=fix(\frac{N}{2})}^{i-1}(4N+1-4fix(\frac{N+1}{2})-1+j\right)\      \\ =A\left(fix\frac{(4i-j+1)}{2},fix\frac{(j+1)}{2}\right)
\end{equation*}
\begin{center}
$j=1,2,\dots,4N+1-4i$\\
$i=fix(\frac{N}{2})+1,fix(\frac{N}{2})+2,\dots,N$\\
\end{center}
The equations above denote the left line map algorithm.To obtain the
right line map,first the image has to be mirrored and then the left
line map is applied to the mirrored image.

This scheme can be extended to rectangular images according to the
size of the image. Using the line map, the encryption of an image is
done as follows: the key for encryption is the number of times the
left and right line map is applied on the image in succession. For
example, a key like `321' means that the image is mapped to another
encrypted image through 3 iterations of the left line map, 2
iterations of the right line map and 1 iteration of the left line
map in that order successively.

\begin{figure}[!hbp]
\centering
\includegraphics[scale=0.5]{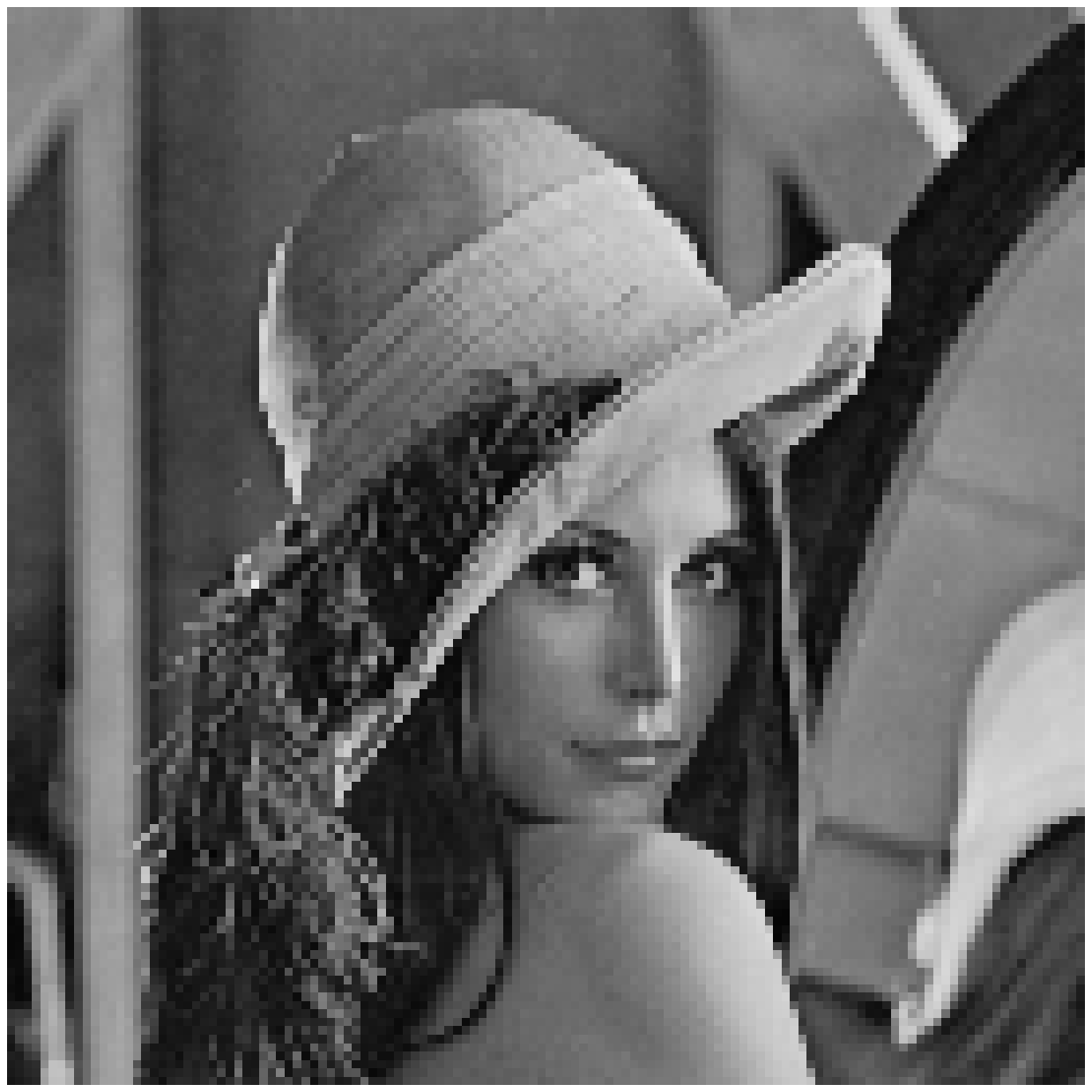}
\hspace{0.5in}
\includegraphics[scale=0.5]{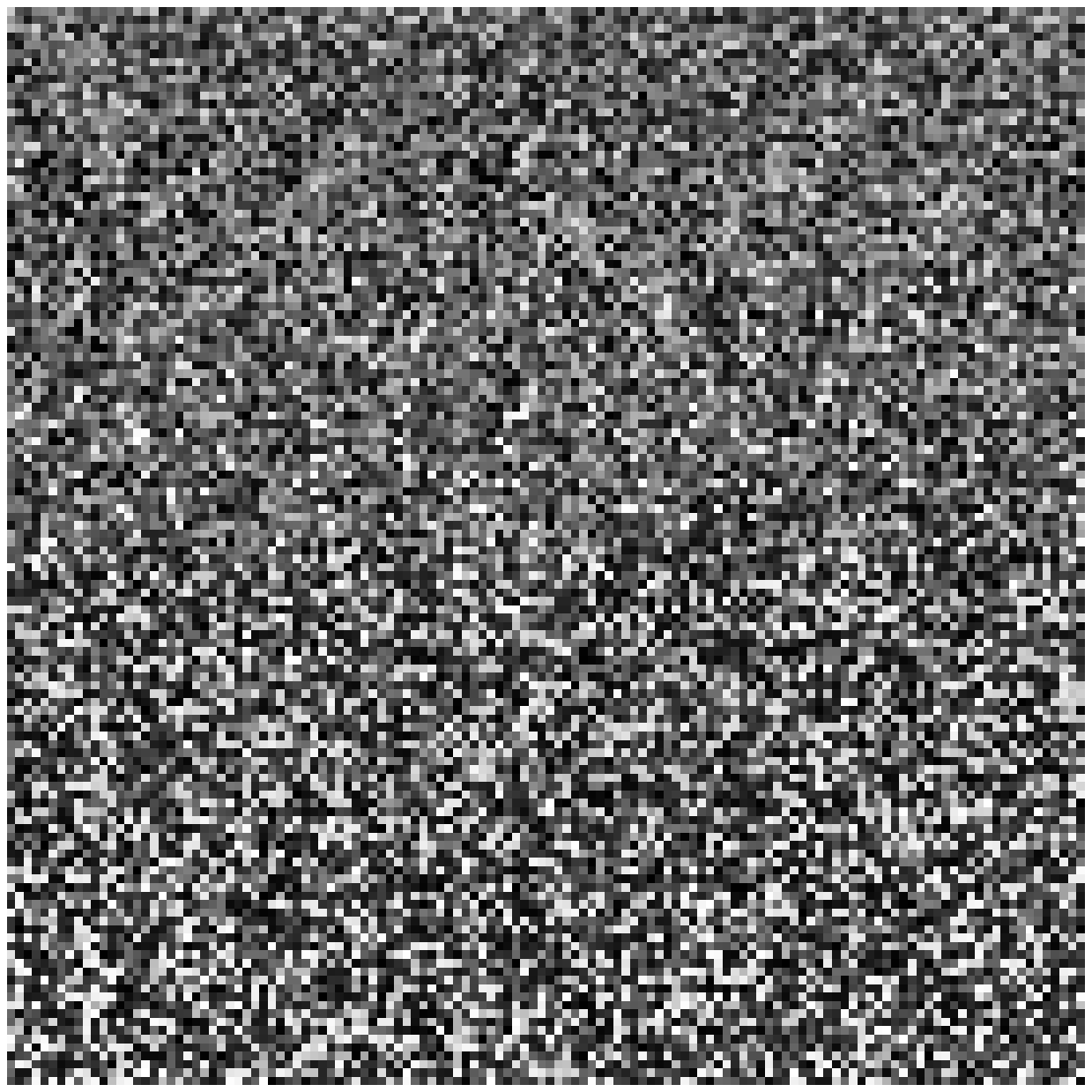}
\caption{(a) Original Image. (b) Encrypted image
(Key:1234567891123456).} \label{fig:fig1}
\end{figure}

For image decryption, we use the inverse line map algorithm, which
is straight away, the reverse of the encryption algorithm. When the
correct key is entered, the original image is obtained without any
loss of information. The line map shows good sensitivity to the
changes in the key. To further enhance the security of the
algorithm, it is proposed that the gray level values substitution is
used together with the line map using XOR operation of the pixel
co-ordinates ($i$ and $j$). This flattens the histogram of the
encrypted image, which resists plaintext attack according
to~\cite{Feng2006}.All the results throughout this paper have been
tested using a $128 \times 128$ Lena image
(Figure~\ref{fig:fig1}(a)).

\section{Weaknesses}
\begin{enumerate}
\item It is claimed in~\cite{Feng2006} that the key space for a key length of 64 bits (16 digits)
is $1.84 \times 10^{19}$. This is actually not possible because the
total number of expressible numbers through 16 digits is $10^{16}$
in a decimal number system. So the key space when key length (K) is
specified/known can be a maximum of only $10^K$.
\item There exists equivalent keys for the encryption operation as defined by the scheme.
As an example, it can be seen that cipher images resulting from keys
4, 103, 301, 202 and 1010101 all reduce to 4 successive iterations
of the line map and are thus equivalent (they produce the same
encrypted image). Hence, even without any attack, the key space is
reduced by a considerable amount.
\item It is also believed that the scheme can have infinite key length.
Practically, since different keys have different times of
encryption, by doing a side channel attack (by recording the time
taken for line map algorithm), the key used can be localized to a
much smaller subset of the key space.
\item The diffusion operation defined by the scheme is weak.
By diffusion we mean that, if a change occurs in a pixel's
gray-level, then the change can cause greater changes in other
pixels resulting in a totally different cipher text after
encryption. Thus, greater the changes caused by diffusion process,
higher is the plaintext sensitivity of the cryptosystem, and the
system is more secure against differential attack. In this scheme,
the XOR of i and j with the plaintext affects only one pixel, which
makes the scheme vulnerable to differential attacks. Moreover, the
diffusion operation is very loosely defined: \textit{``In order to
enhance the security, the gray level values substitution is used
together with the line map using exclusive OR operation of i and
j''}(quoted from Feng~\cite{Feng2006}). Here, the XOR operation if
assumed to be made between the current pixel value and the $i$ and
$j$ coordinates of the pixel, has poor diffusion and affects only
one pixel.
\end{enumerate}

\subsection{Side-Channel attack}

\begin{figure}[!hbp]
\centering
%\includegraphics{chec.eps}
%\hspace{0.1in}
\includegraphics[scale=0.5]{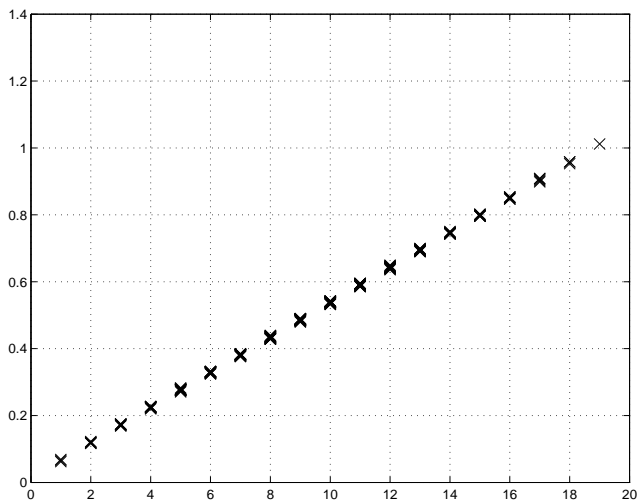}
\caption{ Execution time in seconds vs. sum of iterations (used in
side-channel attack). The experiments were done on a machine running
sempron processor at 1.8 GHz and 256 MB RAM using MATLAB/Dev C++.}
\label{fig:fig2}
\end{figure}

In~\cite{Feng2006}, they claim that since the length of the key has
no limit, its key space can be infinitely large according to
security requirements. Though this feature has been highlighted as
an important advantage of this scheme there are a few glaring
weaknesses in using it. The encryption scheme takes variable time to
encrypt the image for different keys. So a side-channel attack done
by calculating the time taken for varying key lengths greatly
reduces the key space for a brute force attack. We take an example
of key length=3. There are only $10^3$ possible keys of key length
$3$. But among those, a key like `123'(total number of times of
application of either of the line maps is $7$) takes a lot more time
to encrypt/decrypt an image than a key like `111'(total number of
iterations is $3$).This is a grave weakness as it gives away the
number of iterations of the line map involved in the
encryption/decryption process.
The graph (Figure~\ref{fig:fig2}) shows the variation of time taken
for encryption with respect to the total number of iterations due to
the key. From the linear slope, and intercept on the axes, one can
narrow down the key search for brute force attack drastically. We
derive the following results:\\
\begin{enumerate}
\item The upper bound on the size of key space after we have an
estimate of the total number of iterations ($N$) with maximum error
$e$ can be derived from the theory of compositions~\cite{Math} in
number theory (Appendix). (It includes all possible keys of all
possible key lengths for sum of iterations varying between $N-e$ and
$N+e$). When we sum over all possible key lengths ($K$) for this
condition,the key space is given by
$\sum_{n=N-e}^{N+e}\sum_{K=ceil(n/9)}^{N+e}\binom{N-1}{K-1}$

\item For the case where the key length is exactly known (say equal to $K$),
the brute force attack key search is reduced from $10^K$ keys to
$\binom{N-1}{K-1}=\frac{(N-1)!}{(K-1)!(N-K)!}$ keys.
\end{enumerate}

Note that the given estimate of key space does not
 take in to account the reduction due to equivalent keys. In~\cite{Feng2006} a key like $1234567891123456$
 has been used as an example to encrypt images ($N=67$ and $K=16$).
 Even though it is claimed that the key space for such a key length is $1.84 \times 10^{19}$,
 with a knowledge of the key length, it is actually reducible to $\binom{67}{15}$
 which equals $2.683 \times 10^{14}$ -- a reduction in key space by a factor
 of $10^4$. Such a reduction in key space makes the scheme extremely vulnerable against
 the computing facility available at the attacker's disposal (A key
space $K <2^{100} = 10^{30}$ is generally agreed to be insecure in
the cryptography community~\cite{Li2006})

\subsection{Chosen plain text attack}

\begin{figure}[!hbp]
\centering
\includegraphics[scale=0.5]{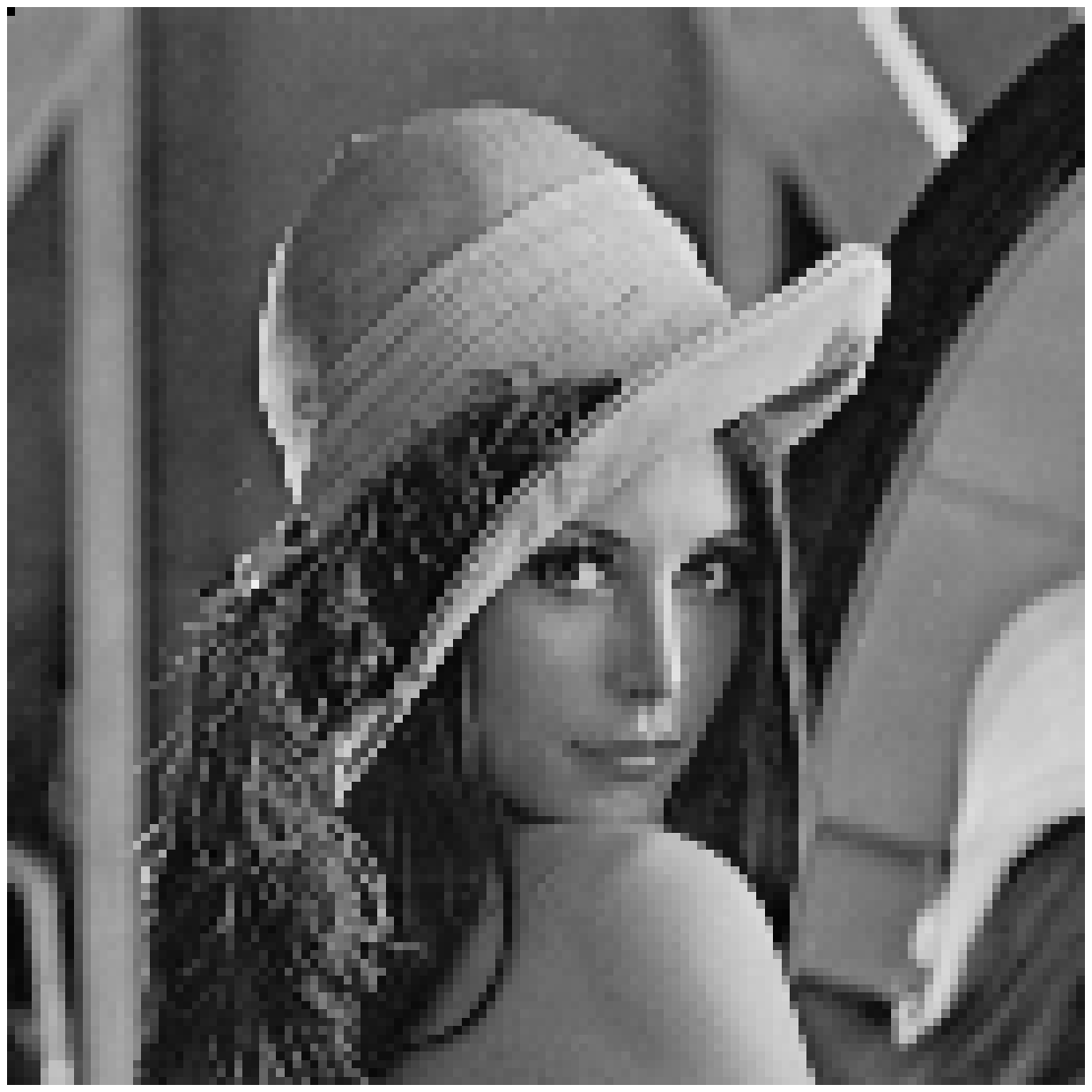}
\hspace{0.5in}
\includegraphics[scale=0.5]{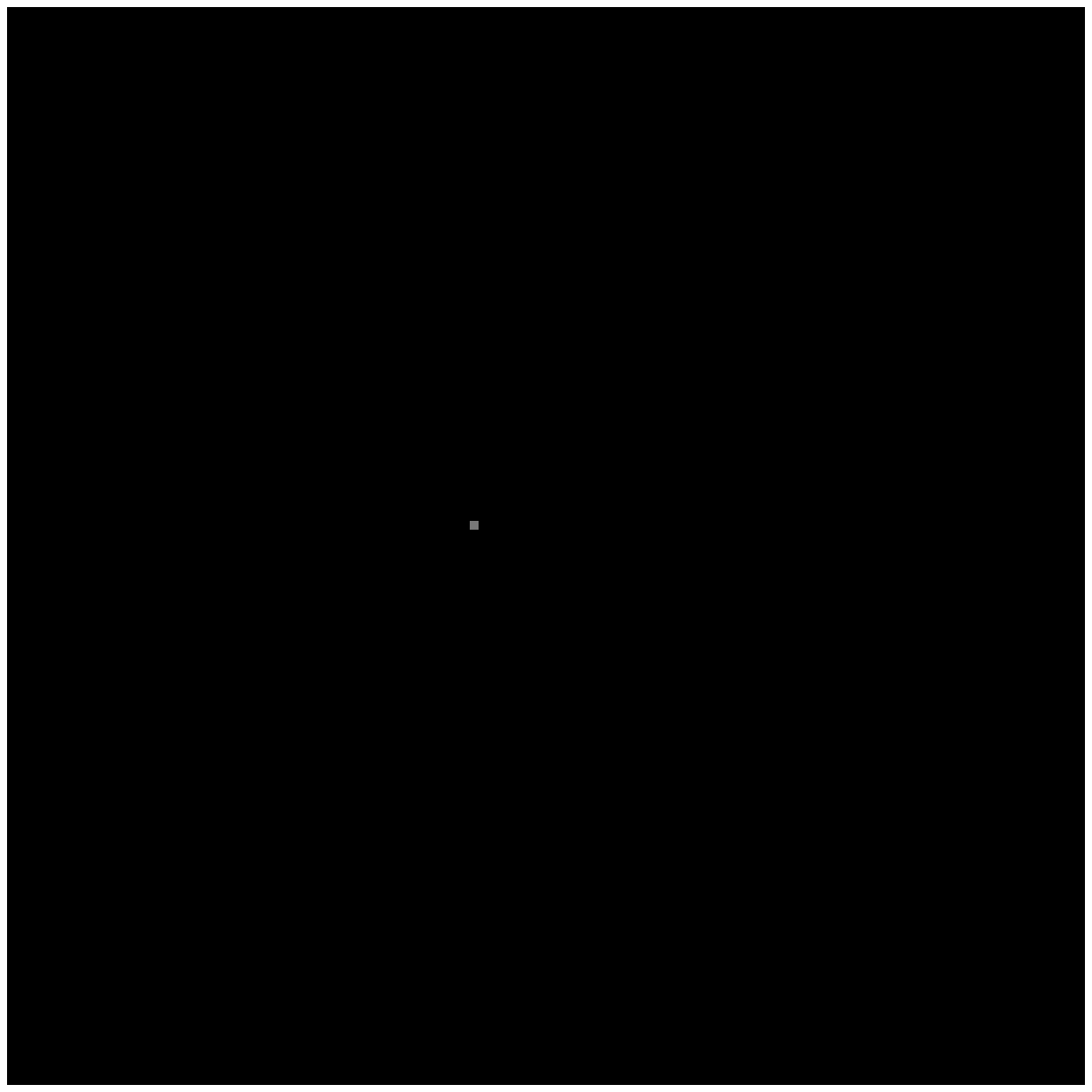}
\caption{(a) Left: Image with only the first pixel different from
the original image. (b) Right: Difference image for
key:1234567891123456. } \label{fig:fig3}
\end{figure}

Even without a brute force attack, the line map scheme is vulnerable
to a known/chosen plaintext attack. According to
Fridrich~\cite{Fridrich1997}, one of the important requirements of a
good cryptosystem is not just sensitivity to the key, but also
sensitivity to small changes in the plaintext. This means that
changing one pixel in the plain-image should result in a completely
different cipher-image. This results when the pixel permutation and
diffusion operations are strong. The line map doesn't possess this
property and hence is vulnerable in this regard. We changed the
first pixel value (137 originally) of the first row of the image to
0. The difference image (difference between original image encrypted
and modified image encrypted) is shown in Figure~\ref{fig:fig3}. We
see that the difference is in one pixel only. This gives a lot of
opportunity to known/chosen plaintext attacks. For instance, if we
choose two images that differ in only one pixel and encrypt it using
an unknown key, we can observe from the difference image how the
different pixel moves after each iteration, from which the key can
be recovered.

\section{Joint Entropy}
Throughout our experiments with encryption and decryption of images,
we have noticed that there exists a need for a good metric to
quantify the amount of randomness in the pixel permutation and
diffusion operations. These operations are necessary for any secure
encryption scheme and a definite way to characterize them will make
comparison of the existing schemes much easier. We propose the Joint
Entropy function of the image as a metric.  Joint Entropy of an
image (assuming a gray image with $2^L$ gray levels) is defined as
the number of bits required to represent every pair of adjacent
pixels of the image at a time (pairs being non-overlapping),
averaged over the entire image. With our test image, there are 256
gray levels, and hence the joint entropy for this image is found as
follows: we find the frequency of occurrence of all possible
combination of gray levels starting from (1,1) to (256,256) and with
this statistical measurement, we obtain the probability of joint
occurrence of any combination of pixel values ($val_1$ and $val_2$).
Joint entropy is then computed as follows:
\begin{equation}
E=\sum_{val_1}\sum_{val_2}p_{val_1val_2}log_{2}(p_{val_1val_2})
\end{equation}

\begin{figure}[!hbp]
\centering
%\includegraphics{chec.eps}
%\hspace{0.1in}
\includegraphics[scale=0.5]{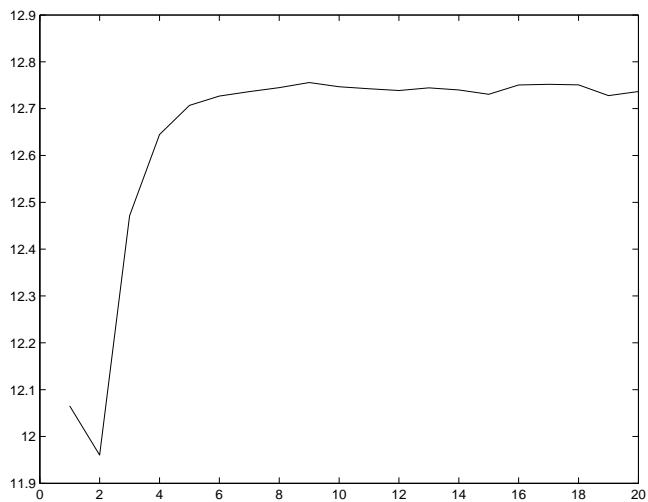}
\caption{Variation of Joint Entropy (in bits) with the successive
iterations of the line map. The first value is the Joint Entropy of
the original image.} \label{fig:fig4}
\end{figure}

\begin{enumerate}
\item Intuitively, entropy is a measure of randomness.
With every iteration of the chaotic map, the pixel arrangement in an
image should become uncorrelated and give a noisy appearance after a
few iterations. This is evident from (Figure~\ref{fig:fig4}) where
there is a steep ascent in the Joint Entropy after a few iterations
of the map (the first value in the graph is the Joint Entropy of the
original image without any iterations of the line map algorithm).
\item Weak keys should be avoided. All keys should produce the same `amount
of randomness'. This means that the attacker should not be able to
distinguish between two ciphertexts by their way of permutation or
diffusion.
\end{enumerate}

For some transformations like cat map, which exhibit periodicity,
the Joint Entropy function can be used to guess the period of the
transformation. We have calculated Joint Entropy here taking in to
account only two adjacent column pixels at a time. But there can be
other ways to do it, like taking two diagonal pixels or two row
pixels or sometimes, more than two pixels at a time to calculate
Joint Entropy, which might yield different results from the ones
obtained.

\section{Conclusions and Future Work}

In this paper, the security of a recently proposed image encryption
scheme has been studied in detail. It is found that there exist some
serious problems with the key scheme, including weak ones and
equivalent ones. The scheme is shown to lack sensitivity to
plaintext albeit possessing sensitivity to changes in the key which
makes it vulnerable to differential attacks. The Joint Entropy
function shows some promise for fulfilling the need for a good
metric to quantify the extent of pixel permutation/diffusion in an
encrypted image. But more research needs to be done on the
relationship between the number of pixels taken in to consideration
while calculating Joint Entropy and the way it affects the Entropy
function, or how much more information it might give about the
encryption scheme.
\section*{Acknowledgements}
Nikhil Balaji would like to express his sincere gratitude to
Department of Electronics and Communication Engineering, National
Institute of Technology Karnataka (NITK) where he is an
undergraduate student in his Final year. Nithin Nagaraj would like
to express his sincere gratitude to Department of Science and
Technology (DST) for funding the Ph.D. program at National Institute
of Advanced Studies (NIAS) which enabled this work. We would like to
thank Prof. Prabhakar G. Vaidya for being a constant source of
inspiration and guidance in our research.

Both the authors express their thanks to Chengqing Li, Department of
Mathematics, Zhejiang University, Hangzhou 310027, China, for useful
comments on the paper.

%
%

%\newpage

\begin{center}
\section*{Appendix}
\end{center}

\subsection*{Composition of numbers}

In number theory, a \textit{composition} of a positive integer $n$
is a way of writing $n$ as a sum of positive integers. It differs
from \textit{partition} in that, the order of summands is taken in
to account while calculating composition, whereas it is not
significant while calculating the number of partitions. For example,
3 has four compositions namely:

\begin{equation*}
\begin{split}
3 & =3\\
  & =2+1\\
  & =1+2\\
  & =1+1+1
\end{split}
\end{equation*}

\subsection*{Number of Compositions}

\emph{Theorem:}\\ Any positive number N can have $2^{N-1}$ compositions.~\cite{Math}\\

\noindent \emph{Proof:}\\ Let us find the number of compositions of
a number $N$. We can write $N$ as the sum of $N$ ones, or ($N-2$)
ones and a 2, ($N-3$) ones and a 3 and so on.  To generalize this
problem, we can say we have to write $N$ as a sum of $K$ parts,
where $K$ varies from 0 to $N$. This is essentially a combinatorial
problem of the number of ways of choosing $K$ objects, given $N$
objects, where $N geq 1$. Such a choice can be made in
$\binom{N-1}{K-1}$ ways. So the total number of ways in which a
composition of the number $N$ can be written is:
\begin{equation}
\sum_{K=1}^{N} \binom{N-1}{K-1}=2^{N-1}
\end{equation}

Also, from combinatorics we know that:

\begin{equation}
\sum_{K=0}^{N} \binom{N}{K}=2^{N}
\end{equation}

which gives a complete proof to the expression for number of
compositions of the number $N$.


\begin{thebibliography}{1}
\bibitem{Mahesh2007}
M. C. Shastry, N. Nagaraj, P. G. Vaidya, ``The B-Exponential Map: A
Generalization of the Logistic Map, and Its Applications In
Generating Pseudo-random Numbers'', arXiv:cs/0607069v2 [cs.CR],
2006.

\bibitem{Feng2006}
Y.Feng, L.Li, F.Huang, ``A Symmetric Image Encryption Approach based
on Line Maps",  1st Int. Symposium on  Systems and Control in
Aerospace and Astronautics 2006(ISSCAA'06), 19-21 Jan. 2006.

\bibitem{Scharinger1998}
J. Scharinger, ``Fast encryption of image data using chaotic
Kolmogorov flows", J. Electronic Imaging, Vol.7, No.2, pp.318-325,
1998.

\bibitem{Fridrich1997}
J. Fridrich J, ``Secure image ciphering based on chaos", Final
report for AFRL Rome, New York, 1997.

\bibitem{Fridrich1998}
J. Fridrich, ``Symmetric ciphers based on two-dimensional chaotic
maps", Int. J. Bifurcation and Chaos, Vol.8, No.6, pp: 1259-1284,
1998.

\bibitem{Lian2005}
S. Lian, J. Sun, Z. Wang, ``Security Analysis of a Chaos-based Image
Encryption Algorithm", Physica A, Elsevier Science, 2005.

\bibitem{Yen1999}
J. Yen, J. Guo, ``A new image encryption algorithm and its VLSI
architecture", In Proc. IEEE Workshop on Signal Processing Systems
(SiPS'99), pp. 430-437, 1999.

\bibitem{Yen2002}
J. Yen, J. Guo, ``Design of a new signal security system", In Proc.
IEEE Int. Symposium on Circuits and Systems (ISCAS'02), Vol. 4, pp.
121-124, 2002.

\bibitem{Li2002}
S. Li, X. Zheng, ``On the security of an image encryption method",
In Proc. IEEE Int. Conference on Image Processing (ICIP'02), Vol. 2,
pp. 925-928, 2002.

\bibitem{Yen2000}
J. Yen, J. Guo, ``A new chaotic key-based design for image
encryption and decryption", In Proc. IEEE Int. Symposium on Circuits
and Systems (ISCAS'00), Vol. 4, pp. 49-52, 2000.

\bibitem{Shujun2002}
S. Li, X. Zheng, ``Cryptanalysis of a chaotic image encryption
method", In Proc. IEEE Int.Symposium on Circuits and Systems
(ISCAS'02), Vol. 2, pp. 708-711, 2002.

\bibitem{Yen1998}
J. Yen, J. Guo, ``A new hierarchical chaotic image encryption
algorithm and its hardware architecture", In Proc. 9th (Taiwan) VLSI
Design/CAD Symposium, 1998.

\bibitem{Guo1999}
J. Guo, J. Yen, J. Yeh, ``The design and realization of a new
hierarchical chaotic image encryption algorithm", In Proc. Int.
Symposium on Communications (ISCOM'99), pp. 210-214, 1999.

\bibitem{ChengYen2000}
J. Yen, J. Guo, ``Efficient hierarchical chaotic image encryption
algorithm and its VLSI realisation", IEEE Proc.-Vision, Image and
Signal Processing, Vol. 147 No.2 pp.167-175, 2000.

\bibitem{SLi2004}
S. Li, C. Li, G. Chen, N. G. Bourbakis, ``A General Cryptanalysis of
Permutation-Only Multimedia Encryption Algorithms",
IACR's Cryptology ePrint Archive, Report 2004/374, 2004.

\bibitem{Mao2004}
Y. Mao, G. Chen, C. K. Chui, ``A symmetric image encryption scheme
based on 3D chaotic cat maps", Chaos, Solitons and Fractals, Vol.
21, No. 3, pp.749-761, 2004.

\bibitem{Kai2005}
K. Wang, Pei, L. Zou, A. Song, Z. He, ``On the security of 3D
Cat map based symmetric image encryption scheme", Physics Letters A,
Vol.343, No. 6, pp.432-439, 2005.

\bibitem{Cli2007}
C. Li, ``On the security of a class of image encryption schemes",
IACR's Cryptology ePrint Archive, Report 2007/339, 2007

\bibitem{Mao2003}
Y. Mao, G. Chen, S. Lian: ``A Novel Fast Image Encryption Scheme
Based on 3D Chaotic Baker Maps", Int. J. Bifurcation and Chaos, June
2003.

%\bibitem{Wolfgang2005}
%K.Kelber, W. Schwarz, "General Design Rules for Chaos-Based Encryption Systems,"
%International Symposium on Nonlinear Theory and its Applications(NOLTA 2005),
%Bruges, Belgium, October 18-21, 2005

\bibitem{Math}
\href{http://www.mathworld.wolfram.com/Composition.html/}{ Wolfram
Mathworld: Composition of functions.}

\bibitem{Li2006}
G. Alvarez, S. Li, ``Some basic cryptographic requirements for
Chaos-based Cryptosystems", Int. J. Bifurcation and Chaos, vol. 16,
no. 8, pp. 2129-2151, 2006.


\end{thebibliography}
\end{document}